 \documentclass[aps,showpacs,amsmath,amssymbd,dvips,showkeys]{revtex4}

\usepackage{graphicx}
\usepackage{color}
\def \be  {\begin{equation}}
\def \ee  {\end{equation}}
\def \ee  {\end{equation}}
\def \bea {\begin{eqnarray}}
\def \eea {\end{eqnarray}}

\newcommand{\nn}{\nonumber}

\begin{document}

\preprint{ECTP-2013-17\hspace*{0.5cm}and\hspace*{0.5cm}WLCAPP-2013-14}

\title{Black Hole Corrections due to Minimal Length and Modified Dispersion Relation}
\author{Abdel~Nasser~TAWFIK\footnote{http://www.atawfik.net/}}
\affiliation{Egyptian Center for Theoretical Physics (ECTP), Modern University for Technology and Information (MTI), 11571 Cairo, Egypt}
\affiliation{World Laboratory for Cosmology And Particle Physics (WLCAPP), Cairo, Egypt}

\author{Abdel~Magied~DIAB}
\affiliation{Egyptian Center for Theoretical Physics (ECTP), Modern University for Technology and Information (MTI), 11571 Cairo, Egypt}
\affiliation{World Laboratory for Cosmology And Particle Physics (WLCAPP), Cairo, Egypt}

\date{\today}

\begin{abstract}
The generalized uncertainty principles (GUP) and modified dispersion relations (MDR) are much like two faces for one coin in research for the phenomenology of quantum gravity which apparently plays an important role in estimating the possible modifications of the black hole thermodynamics and the Friedmann equations. We first reproduce the horizon area for different types of black holes and investigate the quantum corrections to Bekenstein-Hawking entropy (entropy-area law). Based on this, we study further thermodynamical quantities and accordingly the modified Friedmann equation in four-dimensional de Sitter-Schwarzschild, Reissner-N\"{o}rdstrom and Garfinkle-Horowitz-Strominger black holes. In doing this we applied various  quantum gravity approaches. The MDR parameter relative to the GUP one is computed and the properties of the black holes are predicted. This should play an important role in estimating response of quantum gravity to the various metric-types of black holes. We found a considerable change in the thermodynamics quantities. 
We find that the modified entropy of de-Sitter-Schwarzshild and Reissner-N\"{o}rdstrom black holes starts to exist at a finite standard entropy. The Garfinkle-Horowitz-Strominger black hole shows a different entropic property. The modified specific heat due to GUP and MDR approaches vanishes at large standard specific heat, while the corrections due to GUP result in different behaviors. The specific heat of modified de-Sitter-Schwarzshild and Reissner-N\"{o}rdstrom black holes seems to increase, especially at large standard specific heat. In the early case, the black hole cannot exchange heat with the surrounding space. Accordingly, we would predict black hole remnants which may be considered as candidates for dark matter. 

\end{abstract}


\maketitle

\tableofcontents
\makeatletter
\let\toc@pre\relax
\let\toc@post\relax
\makeatother

\section{Introduction}
\label{intro}

The black holes can be characterized by mass (M), electric charge (Q) and angular momentum ($\vec{J}$) \cite{Frolov,Townsend,Padmanabhan}. The uncharged black holes are of de-Sitter-Schwarzschild-type, while the charged ones can be characterized by Reissner -Nordstr\"{o}m metric. There are various quantum gravity approaches designed to study the quantum description of some problems in presence of gravitational fields \cite{Tawfik:2014zca,TD1}. We limit the discussion to the effects of minimal length and/or maximal momentum which are likely applicable at Planck scale which lead to modifications in Heisenberg uncertainty principle in form of quadratic and/or linear terms of momentum. The earlier was  predicted in different theories as string theory, black hole physics and Loop quantum gravity \cite{Amati88,Amati87,Amati90,Maggiore93,Maggiore94,Kempf,Kempf97,
Kempf2000,Kempf93,Kempf94,Kempf95,Kempf96,Scardigli,Scardigli2009}. The latter one was introduced by Doubly Special Relativity (DSR). DSR suggests minimal uncertainty in position and maximum measurable  momentum \cite{DSR,Smolin,Amelino2002a,Amelino2002b,Tawfik:2014zca,TD1}. Furthermore, DSR gives a linear combination of the last two approaches and  amazingly agrees well with the predications of string theory, black hole physics and Loop quantum gravity. Accordingly, a minimum measurable length and a maximum measurable momentum \cite{advplb,Das:2010zf,afa2} are simultaneously likely. This offers a major revision of quantum phenomena \cite{Tawfik:2014zca,TD1,amir,Pedrama,Pedramb}. These approaches have the genetic name, {\it Generalized (gravitational) Uncertainty Principle (GUP)}. Recently, the implication of GUP approaches on different physical systems has been carried out \cite{Tawfik:2014dza,Tawfik:2013uza,Ali:2013ma,Ali:2013ii,Tawfik:2012he,Tawfik:2012hz,Elmashad:2012mq}.

When adding tiny Lorentz-violating terms to a conventional Lagrangian \cite{Glashow:1a,Glashow:1b}, experimental tests can performed by setting upper bounds to the coefficients of these terms, where the velocity of light $c$ should differ from the maximum attainable velocity of a material body. This small adjustment of $c$ leads to a modification in the energy-momentum relation and possible $\delta \textit{v}$ \cite{Glashow:1a,Glashow:1b,Glashow:2,Glashow:3,Amelino98} so that the vacuum dispersion relation becomes sensitive to the type of quantum gravity effect. In additional to that, the possibility that the relation connecting energy and momentum in Special Relativity (SR) may be modified at Planck scale because of the threshold anomalies of ultra-high energy cosmic ray (UHECR). This enters the literature as {\it Modified Dispersion Relations (MDRs)} \cite{Amelino98,Amelino2001,Amelino2001b,Amelino2004,Amelino2004b,
Amelino2006,Nozari2006,Aloisio,Jacobson,Nozari} and can provide new sensitive tests for SR. Successful searches would reveal a surprising connection between particle physics cosmology \cite{Glashow:1a,Glashow:1b,Glashow:2,Glashow:3}. The modifications of energy-momentum conservations laws of interaction such as pionphotoproduction by inelastic collisions of cosmic-ray nucleons with the cosmic microwave background and higher energy photon propagating in the intergalactic medium which can suffer inelastic impacts with photons in the Infrared background resulting in the production of electron-positron pairs \cite{Y. J. Ng:1995,Y. J. Ng:2000} are examples about MDR.

The finding that the black holes should follow a well-defined entropy-area law from quantum geometry approach shows that $S_{BH}=A/4 \ell _{p}^{2} $ \cite{Bekensteina,Bekensteinb,Bekensteinc,Hawking}, where $A$ is the cross-sectional area of black hole horizon and  $\ell_{p}=\sqrt{\hbar G/c^{3}}$ is the Planck length. Therefore, the connection between the {\it geometric} entropy (and indirectly all other thermodynamic quantities) of black hole and the Planck length (and through it the possible modifications through GUP and/or MDR is apparent. The systematic study of the black hole radiation and the correction due to entropy/area relation gain the attention of theoretical physicists. For instance, there are nowadays many methods for calculating Hawking radiation \cite{Parikh,Dehghani,Srinivasan,Jianga,Jiangb,Barun}. Nevertheless, all results show that the black hole radiation is very close to the black body spectrum \cite{Perlmutter,Caldwell}. This conclusion raised a very difficult question whether the information is conserved through the black hole evaporation process? \cite{Perlmutter,Caldwell}.  The study of thermodynamical properties of black holes in space-times is therefore a very relevant and original task. For instance, based on recent observations of supernova, the cosmological constant may be positive \cite{Perlmutter,Caldwell}!  

In the present work, all possible interrelations shall be calculated by means of  the quantum gravity approaches. Through the comparison of the corrected results obtained from these alternative approaches, it can be shown that a suitable choice of the expansion coefficients in the modified dispersion relations leads to the same results in GUP approach. We first introduce in section \ref{analysis} the different quantum gravity approaches which  shall be utilized  in studying the minimal length and the modification of the energy-momentum relation in UHECR, for instance,. In section \ref{Schwarzschild}, we compute the modified thermodynamical quantities and the modified Friedmann equation in four dimensional de Sitter-Schwarzschild black hole based on GUP and MDR approaches. Section \ref{Reissner} shall be devoted to four dimensional Reissner-N\"{o}rdstrom black hole. The corrections to Garfinkle-Horowitz-Strominger black hole is elaborated in section \ref{Garfinkle}. A comparison between the three types of black holes is given in section \ref{compare}.

\section{Approaches}
\label{analysis}

\subsection{Energy-Area Law of Black Hole}

The entropy-area relation is related to thermodynamical properties of black holes and gained a remarkable attention among physicists \cite{Bekensteina,Bekensteinb,Bekensteinc,Hawking,Parikh,Dehghani,Barun,Carroll,Kourosh,Kourosh:B,Zhu,Reissner}. On the other hand, this reflects an intrinsic property of the black holes. Nevertheless, the gravitational effect on the evolution of the Universe was neglected in many models \cite{Carroll,Srinivasan,Parikh,Bekensteina,Bekensteinb,Bekensteinc,Y. J. Ng:2000}. Over the last decades, this was assumed as a quantum geometric correction in the thermodynamical properties including the characteristic Hawking temperature and entropy \cite{Dehghani}. This should not prevent the combination of quantum gravity with the black hole physics. It has been shown that the correction of quantum geometry with different types of black holes is not just according to the entropy-area law \cite{Bekensteina,Bekensteinb,Bekensteinc} but also according the differences between black hole thermodynamics and FLRW Universes due to different approaches describing  the quantum gravity.

For reference about the entropy-area relation in different black holes under the effects of quadratic GUP, the readers are advised to consult Ref. \cite{Dehghani,Kourosh}. This shows that the quadratic GUP and MDR effects on quantum geometry and on the entropy-area relation \cite{Dehghani,Kourosh}. In the present work, we summarize the effects of minimal length and higher order GUPs and MDRs relations using a rather simple technique.  All possible thermodynamics and modifications in FLRW equation were eliminated for most major black holes such as de Sitter-Schwarzschild and Reissner-Nordstrom black holes and for the charged dilation black hole (Garfinkle-Horowitz-Strominger).  Both modified and unmodified cases are compared with each other.  This offers the possibility to reach some conclusive conclusion. 
\begin{itemize}
\item Firstly, the GUP which is estimated through {\it gedanken} experiments proceeds by observing photons scattered by the studied black hole are based on predications of the MDR relation. The correction of $c$ at the range of UHECR and gamma rays events at the Large Hadron Collider (LHC) energies could predict the existence of black hole remnants, which might be considered as candidates for dark matter and can be used to investigate the possibilities of finding black holes in LHC. 
\item Secondly, we revise the physical concept of the higher order GUP in the momentum space and give the maximum momentum measurable at the uncertainty of minimal length, which apparently agrees with the predication of Double special relativity (DSR) \cite{Smolin} and Hilbert space \cite{Tawfik:2014zca,TD1} and other results estimated from the different behaviors given in section (\ref{compare}).
\end{itemize}

\subsection{GUP: Minimal Length Uncertainty}

Recently, the GUP has been the subject of much interesting works \cite{Tawfik:2014zca,TD1}. It is shown that the uncertainty principle should be modified at the Planck scale regime of $10^{39}$ GeV. \cite{Tawfik:2014zca,TD1,Amati88,Amati87,Amati90,Maggiore93,Maggiore94,Kempf,Kempf97,
Kempf2000,Kempf93,Kempf94,Kempf95,Kempf96,Scardigli,Scardigli2009}
\bea 
\label{KMM}
\Delta x\, \Delta p &\geq & \frac{\hbar}{2}\left(1+ \alpha ^{2} \, (\Delta p)^{2}\right), 
\eea
where $\alpha=\alpha_0 \sqrt{1/(M_p\, c^2) }= \alpha_0 l_p /\hbar $ is a constant coefficient referring to the gravitational effects on Heisenberg uncertainty principle.
The Heisenberg algebra of GUP  \cite{Tawfik:2014zca,TD1,Kempf,Kempf97,Kempf2000,Kempf93,Kempf94,Kempf95,Kempf96} is given as  
\bea 
\left[\hat{x},\, \hat{p}\right] &=& i\, \hbar\, \left(1+ \alpha ^{2}  p^{2}\right). \label{KMM1}
\eea
This can be represented in momentum space wave functions $\phi(p)=\langle p| \phi (p) \rangle $ and $\partial_{p}=i \hbar \partial/\partial {x}$ \cite{Tawfik:2014zca,TD1} .
\bea 
\hat{p}\, \cdot \phi(p) &=& p_0\, \phi(p), \label{eq:app} \nn \\
\hat{x}\, \cdot \phi(p) &=& i\, \hbar\, \left(1+ \alpha ^{2}  p^{2}\right) \partial_{p}\, \phi(p). \label{eq:app1}
\eea
Again, this allows to have non-commutative geometry of the spacetime \cite{Tawfik:2014zca,TD1}, where
\bea 
\left[\hat{p}_{i},\hat{p}_{j}\right]&=&0,\nn \\
\left[\hat{x}_{i},\hat{x}_{j}\right]&=& -\, 2\, i\, \hbar\, \alpha ^{2}\, \left( 1+ \alpha ^{2}\, \vec{\textbf{p}}^{2} \right)\hat{L}_{ij}.
\eea
and the representation of the generators of rotations on momentum wavefunctions  as 
\be 
\hat{L}_{ij}\, \phi(p) = -\, \, i\, \hbar\, \left( p_{i}\, \partial_{p_{j}} - p_{j}\, \partial_{p_{i}}\right)\, \phi(p).
\ee

To compute a minimum uncertainty in position, Eq.  (\ref{KMM1}) and the relation $(\Delta\, p)^{2}=\langle p^{2}\rangle -\langle p\rangle^{2}$ can be utilized \cite{Kempf95}
\bea \label{qua}
\Delta x\, \Delta p &\geq & \frac{\hbar}{2}\left(1+\alpha ^{2} (\Delta p)^{2} +\alpha ^{2} \langle p \rangle ^{2} \right).
\eea
The last expression can be rewritten as a second order equation for $\Delta p$. Then, the solutions for $\Delta p$  \cite{Kempf95}
\bea
\Delta p &=& \left(\frac{\Delta x}{\hbar\,\alpha ^{2}}\right) \pm \sqrt{\left(\frac{\Delta\, x}{\hbar\,  \alpha ^{2}}\right)^{2}-\frac{1}{\alpha ^{2}}-\langle p \rangle^{2}}.
\eea
A minimum value $\Delta x$ can be deduced $\Delta x_{min} (\langle p \rangle) = \hbar \alpha  \sqrt{1+\alpha ^{2} \langle p \rangle ^{2}}$. Therefore, the absolutely smallest uncertainty in position, where $\langle p \rangle =0$, reads 
$\Delta x_{0}=\hbar \alpha$,
where $\alpha ^{2}$ is compatible to $\beta$; GUP parameter \cite{Kempf95}. In natural units, $c=k=\hbar=G=1$. From Eq.  (\ref{KMM}), we have
\bea
\label{quadratic1}
\label{PQGUP}
\Delta p \geq \frac{1}{\Delta x} \lbrace 1+\frac{\alpha ^{2}}{(\Delta x)^{2}}+2 \frac{\alpha ^{4}}{(\Delta x)^{4}}+\cdots \rbrace.
\eea

\subsection{Higher order GUP: Minimal Length Uncertainty and Maximum Measurable Momentum}

The commutator relations \cite{advplb,Das:2010zf,afa2}, which are consistent with string theory, black holes physics and DSR leads to
 \bea
\left[\hat{x}_{i}, \hat{p}_{j} \right]=i \hbar \left[ \delta_{i j} -\alpha \left( p \delta _{i j} +\frac{p_{i} p_{j}}{p} \right)+\alpha ^{2} \left(p^{2} \delta_{ij} +3 p_{i} p_{j} \right)\right], \label{ali1}
\eea
implying minimal length uncertainty and maximum measurable momentum by working with the convenient representation of the commutation relations on momentum space wavefunctions in momentum-space \cite{Tawfik:2014zca,TD1,amir}. The momentum $\textbf{p}$ and position $\textbf{x}$ operators are given as  
\bea 
\hat{x}_{i}\, \phi (p) &=& i \hbar (1 -\alpha \vec{p}_{0} +2\, \alpha^{2}\, p_{0} ^{2})\partial _p\, \phi (p), \label{amm1}\nn \\
\hat{p}_{j}\, \phi (p) &=&  p_{0 j}\, \phi (p), \label{amm2}
\eea
where the minimal length uncertainty \cite{advplb,Das:2010zf,afa2}  and maximum measurable momentum \cite{Tawfik:2014zca,TD1,amir}, respectively, read
\bea 
\Delta x &\geq & (\Delta x)_{min} \approx \hbar \alpha, \nn \\
p_{max} &\approx & \frac{1}{4 \alpha}.
\eea
The maximum measurable momentum agrees with the value which was obtained in the DSR \cite{DSR,Tawfik:2014zca,TD1}. Then, in one dimension in natural units the uncertainty relation  reads \cite{advplb,Das:2010zf,afa2}
\bea 
\Delta x\, \Delta p &\geq & \frac{\hbar}{2} \left(1-2\, \alpha\, \Delta p +4\, \alpha^{2} \Delta p^{2}  \right). \label{ali3}
\eea
This representation of the operators product satisfies the non-commutative geometry of spacetime \cite{amir},
\bea
\left[\hat{p}_{i},\hat{p}_{j}\right]&=&0,\nn \\
\left[\hat{x}_{i},\, \hat{x}_{j} \right] &=& - i\, \hbar\, \alpha\, \left(4\, \alpha - \frac{1}{P}\right)\, \left(1 - \alpha\, p_{0} +2 \alpha^{2}\, \vec{p_{0}}^{2}\right)\; \hat{L_{i j}}. 
\eea

The generators of rotations ${\bf L}_{i j}$ can be expressed in terms of the position and momentum operators \cite{amir}
\bea 
\hat{L}_{ij} \phi(p) &=& -\, i\, \hbar\,  \left( p_{i}\, \partial_{p_{j}} - p_{j}\, \partial_{p_{i}}\right)\, \phi(p).
\eea
From Eq.  (\ref{ali3}) in $\mathcal{O(\alpha)}$, we have 
\bea 
\label{LGup}
\Delta p \geq \frac{1}{\Delta x}\left(\frac{1}{1+\frac{2 \alpha}{\Delta x}}\right).
\eea

\subsection{MDR: Modified Energy-Momentum Relation}
\label{MDRALL}

The energy-momentum tensor characterizing SR, $p^{2} = E^{2} - m^{2}$, may be modified in the Planck scale regime. Anomalies in ultra-high cosmic rays and $\gamma$-TeV events at LHC, may be explained by modification in the dispersion relation   \cite{Amelino98,Amelino2001,Amelino2001b,Amelino2004,Amelino2004b,
Amelino2006,Nozari2006,Aloisio,Jacobson,Nozari}.
\begin{equation}\label{1}
\overrightarrow{p}.\overrightarrow{p}=(\overrightarrow{p})^2=f(E,m;\ell_{p})\simeq E^2-\mu^2+\alpha_1 \ell_{p}^{2}
E^4+\alpha_2 \ell_{p}^4 E^6+{\cal O}\left(\ell_{p}^6 E^8\right), 
\end{equation}
where $f$ is the function of the exact dispersion relation.The applicability of a Taylor-series expansion for $E\ll
1/\ell_{p}$ is assumed. The coefficients $\alpha_i$ may take different values in different quantum gravity approaches. Here, $m$ is the rest mass of the particle and the mass parameter $\mu$ on the right hand side is directly related to the rest energy, but $\mu\neq m$, when  $\alpha_i$s do not vanish.  Differentiation of Eq.  (\ref{1}) leads to
\bea
dp &\approx & d E\, \left[1-\frac{3}{2} \alpha _{1} \ell_{p}^{2}E^{2}+\left(\frac{5 \alpha _{2}}{2}- \frac{5}{8}\alpha_{1} ^{2}\right) \ell_{p}^{4}E^{4}\right], \\
dE &\approx & d p \left[1-\frac{3}{2} \alpha _{1} \ell_{p}^{2}E^{2}+\left(\frac{5 \alpha _{2}}{2}- \frac{23}{8}\alpha _{1} ^{2}\right) \ell_{p}^{4}E^{4}\right]. \label{8}
\eea
One can obtain the corresponding generalized relation in $\mathcal{O}(E^{4}) $ as follows. In natural units,  $\ell_{p}$ can be omitted, then
\bea
\label{energymdr}
E \Delta x \geq 1- \frac{3 \alpha _{1}}{2} \frac{\ell_{p}^{2}}{(\Delta x)^{2}}- \left(\frac{5 \alpha _{2}}{2}-\frac{23 \alpha _{1}^{2}}{8}\right)\frac{\ell_{p}^{4}}{(\Delta x)^{4}}.
\eea


\section{de Sitter-Schwarzschild  Black Hole}
\label{Schwarzschild}

In GR, the unique spherically symmetric vacuum solution is the Schwarzschild metric, which in spherical coordinates ($t,r,\theta ,\phi$) \cite{Carroll}, reads
\bea
ds^{2}=-\left(1-\frac{2M}{r}\right) dt^{2}+\left(1-\frac{2M}{r}\right)^{-1} dr^{2}+r^{2}d\Omega ^{2}. \label{met}
\eea
Then, the horizon area $A$ and entropy $S$, energy density $\rho$, Hawking radiation temperature $T$, and specific heat capacity at constant volume $C_{v}$, respectively, reads 
\bea
A&=& 4 \pi r_{H}^{2}=16 \pi M^{2}, \label{1} \\
S&=&4 \pi M^{2}, \label{2} \\
\rho &=& 
\frac{3}{2 A}, \label{3} \\
T &=& \frac{1}{4 \pi r_{H}}=\frac{1}{8 \pi M}, \label{4} \\
C_{v} &=& 12\, \pi \, T. \label{5}
\eea
The corrections to these five intensive and extensive thermodynamic quantities shall be estimated in the sections that follow. Accordingly, we introduce the possible modifications in Friedmann equation. We first utilize quadratic GUP in section \ref{sec:bhQGUP}.  The modifications due to linear GUP shall be elaborated in section \ref{sec:bhLGUP}. 

\subsection{Quadratic GUP and Black Hole Thermodynamics}
\label{sec:bhQGUP}


According to the metric equation of de Sitter-Schwarzschild black hole, Eq.  (\ref{met}), and the uncertainty of momentum, Eq.   (\ref{PQGUP}), we have 
\bea 
d A &=& 32 \pi M \frac{1}{\Delta x}, \\
d A_{GUP} &=& dA  \frac{1}{\Delta x} \left[1+\frac{\alpha ^{2}}{(\Delta x)^{2}}+2 \frac{\alpha ^{4}}{(\Delta x)^{4}}+\cdots \right].
\eea
If we utilize the notations introduced in Ref. \cite{Medved}, $\Delta x= 2 r_H = \sqrt{A/\pi}$.
The modified horizon area and entropy of de Sitter-Schwarzschild black hole, respectively, read
\bea
A_{GUP} &=& A+\alpha ^{2} \pi \ln{A}-\, 2\,(\alpha ^{2} \pi)^{2} \frac{1}{A}+\cdots. \\
S_{GUP} &=& \frac{A}{4}+ \frac{\alpha ^{2} \pi}{4} \ln{A}-\frac{(\alpha ^{2}\pi)^{2}}{2} \frac{1}{A}+\cdots+C. \label{SA}
\eea
From the Bekenstein-Hawking area law $s=A/4$ \cite{Bekensteina,Bekensteinb,Bekensteinc,Hawking}, the modified entropy of de Sitter-Schwarzschild black hole from quadratic GUP approach can be expressed in term in unmodified entropy $s$, 
\bea
S_{GUP}= s +  \frac{\alpha ^{2} \pi}{4} \ln{s} - \frac{\pi ^{2} \alpha ^{4}}{8}\, \frac{1}{s}+\cdots +C, \label{sss}
\eea
where $C$ is an arbitrary constant


The black hole energy density can computed as follows \cite{Kourosh,Kourosh:B,Zhu}.
 \bea
 \frac{8 \pi }{3} \rho = - \pi \int S^{'}(A) \left(\frac{A}{4}\right)^{-2} dA.  \label{rhoi}
 \eea
The modified energy density of de Sitter-Schwarzschild black hole from quadratic GUP approach is given as
\bea
\rho_{GUP} &=&\frac{3}{2 A}+ \frac{3 \alpha ^{2} \pi}{4 A^{2}}+\frac{(\alpha ^{2} \pi)^{2}}{4 A^3}+\cdots+C.
\eea
By using Eq.  (\ref{3}), we get 
\bea
\label{rhoq}
\rho_{GUP}=\rho + \frac{\alpha ^{2} \pi}{3} \rho ^{2} +2 \frac{(\alpha ^{2} \pi)^2}{27} \rho ^{3}+\cdots+C,
\eea
where $C$ is an arbitrary constant


From Eq.  (\ref{SA}) and $dM = T dS$, the modified (generalized) Hawking temperature is given as
\bea
T_{GUP}&=&\frac{1}{8 \pi M } \left\lbrace  1+ \frac{\alpha^{2} \pi}{16  M}(1+\frac{\alpha^{2} \pi}{ 8 M^{2}})\right\rbrace ^{-1}=T \, \left\lbrace 1\,-\frac{\alpha^{2} \pi}{2}\, T -4\,(\alpha^{2} \pi )^{4}\, T^{3}\right\rbrace .
\eea


The correction of the specific heat capacity corresponding to the Hawking radiation temperature reads
\bea
C_{v_{GUP}} &=& 12\, \pi \, T\,+\,48\, \pi ^2  \alpha ^2\, \pi \,  T^3\,+\,96 (  \alpha ^2 \pi ^2 )^2\, \pi \,T^5 =
 C_{v}+ \frac{\alpha ^{2}}{72}\, \left\lbrace  2\, C_{v}^{3}\,+ \frac{\alpha ^{2}}{36} \,C_{v}^{5} \right\rbrace .
\eea


Using the corrected entropy-area relation, one can derive the modified Friedmann equations. For an apparent horizon of Friedmann-Lemaitre-Robertson-Walker (FLRW) Universe, Freidmann second equation is given as \cite{Kourosh,Kourosh:B,Zhu}
\bea
\label{fred}
\left(\dot{H} - \frac{K}{a^{2}} \right) S^{'}(A)\, =-4 \pi (\rho +p),
\eea
where $S^{'}(A) \equiv d S(A)/dA.$ From Eqs. (\ref{SA}), (\ref{sss}) and (\ref{rhoq}), the modified Friedmann second equation  reads
\bea
\left(\dot{H} - \frac{K}{a^{2}} \right)\left[1+\frac{\alpha ^{2} \pi }{4s} +\frac{(\alpha ^{2} \pi)^{2}}{32 s^{2}}+\cdots \right] &=&-16\pi \left[p+\rho + \frac{\alpha ^{2} \pi}{3} \rho ^{2} +2 \frac{(\alpha ^{2} \pi)^2}{27} \rho ^{3}+\cdots \right].
\eea

\subsection{Linear GUP and Black Hole Thermodynamics}
\label{sec:bhLGUP}

In $\mathcal{O}(\alpha)$ the uncertainty in momentum, Eq.  (\ref{LGup}), was satisfied \cite{advplb}. We distinguish the modified quantities by an index $DSR-GUP$. The horizon area can be calculated in the same way as done in previous section
\bea
dA_{DSR-GUP} &=& \left(\frac{1}{1+\frac{2 \alpha}{\Delta x}}\right) dA, \\
A_{DSR-GUP} &=& A\, -4\, \alpha \sqrt{\pi} \sqrt{A}+8\, \alpha ^{2} \pi \, \ln{\left(\sqrt{\frac{A}{\pi}}+2 \alpha\right)}.
\eea

Using the Bekenstein-Hawking entropy-area law  \cite{Bekensteina,Bekensteinb,Bekensteinc,Hawking}, we obtain
\bea
S_{DSR-GUP} &=& \frac{A}{4}- 2\alpha \sqrt{\pi} \sqrt{\frac{A}{4}}+2\alpha ^{2} \pi \ln{\frac{A}{4}}+\cdots+C, \label{SaAA} \\
S_{DSR-GUP}&=& s -2 \alpha \sqrt{\pi} \sqrt{s}+2\alpha ^{2} \pi \ln{s}+\cdots+C,
\eea
where $C$ is an arbitrary constant.

Differentiating the entropy, Eq. (\ref{SA}), and using Eq.  (\ref{rhoi}), then the modified energy density of the de Sitter-Schwarzschild black hole based the linear GUP reads
\bea
\rho _{DSR-GUP}= \rho -\frac{4}{3}\, \left(\sqrt{\frac{2}{3}} \alpha \sqrt{\pi}\right) \rho ^{3/2} +\frac{1}{2}\left(\sqrt{\frac{2}{3}}\alpha \sqrt{\pi}\right)^{2} \rho ^{2} +\cdots +C,
\eea
where $C$ is an arbitrary constant

In linear GUP approach, the characteristic Hawking temperature  is given as
\bea
T_{DSR-GUP} &=& \frac{1}{8 \pi M}\, \left\lbrace  1-\left(\frac{\alpha}{2}\right)\, \frac{1}{M}+\left(\frac{\alpha}{2}\right)^{2}\, \frac{1}{M^{2}}+\cdots \right\rbrace , \\
T_{DSR-GUP} &=& T\, \left\lbrace  1+4\, \pi \alpha \, T\,-(4 \pi \alpha)^{2}\, T^{2}\,+\cdots \right\rbrace .
\eea

Based on the same GUP approach, the corresponding specific heat capacity can estimated as
\bea
 C_{v{DSR-GUP}} &=& 12 \, \pi  T-48\, \pi ^2 \alpha\, \,T^2 \,+48\, \pi ^3 \, \alpha ^2\, T^3, \\
 C_{v{DSR-GUP}} &=& C_{v}-\, \frac{1}{3} \alpha \,C_{v}^{2} +\,\frac{\alpha ^{2}}{36 }\, C_{v}^{3}.
 \eea
 

At apparent horizon of FLRW Universe, the Friedmann second equation gets modifications due to the linear GUP approach 
\bea
\left(\dot{H}+\frac{K}{a^{2}}\right)\left[1-\frac{\alpha \sqrt{\pi}}{\sqrt{s}}+\left(\frac{\alpha \sqrt{\pi}}{\sqrt{s}}\right)^{2}+\cdots \right]
&=&-16\pi \,\left[p+ \rho -\frac{4}{3}\sqrt{\frac{2}{3}} \alpha \sqrt{\pi}\, \rho ^{3/2} +\frac{1}{2}(\sqrt{\frac{2}{3}}\alpha \sqrt{\pi} )^{2}\, \rho ^{2} +\cdots \right]. \hspace*{8mm}
\eea

\subsection{MDRs and Black Hole Thermodynamics}

If a quantum particle with energy $E$ and size $\ell$ is absorbed into a black hole and $\ell \approx \Delta x$, then minimum increase of area of black-hole $\Delta A\, =\,4\, E\, \Delta x \, \ln{2}$ and
\bea
 \frac{dS}{dA}\approx \frac{\Delta S_{min}}{ \Delta A_{min}} \approx \frac{\ln{2}}{4 E \Delta x \ln{2}}.  \label{smdr}
\eea
 
Accordingly, we get 
\bea
dS_{MDR} &=& \frac{1}{4}\left(1 -\, \frac{3 \alpha _{1} \pi}{2}\, \frac{1}{A}-\, \frac{5 \pi ^{2}}{8}\left(\alpha _{1}^{2} - 4\alpha _{2}  \right) \frac{1}{A^{2}}\right) dA, \\
 S_{MDR} &=&\, \frac{A}{4}+\, \frac{3 \alpha _{1} \pi}{8} \ln{\frac{A}{4}} + \,  \frac{5 \pi ^{2}}{8} \left(\alpha _{1}^{2} - 4 \alpha _{2}  \right) \frac{4}{A}= s+\, \frac{3 \alpha _{1} \pi}{8} \ln{s} + \frac{5 \pi ^{2}}{8} \left(\alpha _{1}^{2} -4  \alpha _{2}  \right) \frac{1}{s}+\cdots +C, \label{sMD}
\eea
where $C$ is an arbitrary constant. When comparing Eq.   (\ref{sMD}) with Eq.    (\ref{sss}),  the values of $\alpha _{1} $ and $\alpha _{2}$ corresponding to the GUP parameter $\alpha$ can be estimated
\bea
\alpha_{1} &=&\frac{2}{3} \alpha,\\ \nn
\alpha_{2} &=&\frac{41}{45}\alpha^{2}.
\eea
This means that both parameters are compatible with each other and should have the same effect.

From the Bekenstein-Hawking entropy, the modified horizon area of black hole can be computed  
\bea 
A_{MDR}=\, A+\, \frac{3 \alpha _{1} \pi}{8} \ln{A}+\,  \frac{5 \pi ^{2}}{8} \left(\alpha _{1}^{2} -4 \alpha _{2}  \right) \frac{1}{A}.
\eea

From the derivation of Eqs. (\ref{sMD}) and  (\ref{rhoi}), and utilization of Eq. (\ref{3}), the he energy density can be calculated
\bea 
\rho _{MDR}=\, \rho \,- \frac{\alpha _{1} \pi}{2}\, \rho ^{2} -\frac{5 \pi ^{2}}{12}\,\left(\alpha _{1}^{2} - 4\alpha _{2}  \right) \rho ^{3}+C,
\eea
where $C$ is an arbitrary constant.

The modified Hawking radiation reads 
\bea
T_{MDR} &=& \frac{1}{8 \pi M} \left\lbrace 1+\, \frac{3 \alpha _{1}}{32 M^{2}} - \frac{5 \pi}{1024} \left(\alpha _{1}^{2} - 4 \alpha _{2}  \right)\frac{1}{4M^{4}}\right\rbrace  ^{-1}, \\ 
T_{MDR} &=& T \left\lbrace  1-\, \frac{3 \alpha _{1} \pi ^{2}}{2}\, T^{2} + \, \frac{5 \pi ^{5}}{16}\left(\alpha _{1}^{2} - 4 \alpha _{2}  \right) T^{4}\right\rbrace .
\eea

The geometric correction of specific heat capacity relative to the Hawking temperature is given as 
\bea
C_{v_{MDR}} &=& 12 \pi \, T - 72 \pi ^{3} \alpha_{1} T^{3} - 540 \, \pi ^{5} \left(\alpha _{1}^{2} -4 \alpha _{2}  \right)\, T^{5}, \\
C_{v_{MDR}} &=& C_{v} - \frac{\alpha_{1}}{24} C_{v}^{3}+\frac{5}{2304}\,\left(\alpha _{1}^{2} - 4\alpha _{2}  \right)\,C_{v}^{5}.
\eea


For an apparent horizon of FLRW Universe, we can obtain the Friedmann second equation based on the MDRs
\bea
\left(\dot{H} - \frac{K}{a^{2}} \right)\left[1-\frac{3 \alpha _{1} \pi}{8s} - \frac{5 \pi ^{2}}{64s} \left(\alpha _{1}^{2} - 4\alpha _{2}  \right) +\cdots+\right] 
&=&-16 \pi  \left[p+\rho - \frac{\alpha_{1} \pi}{2} \rho^{2} - \frac{5 \pi ^{2}}{12}\left(\alpha _{1}^{2} - 4\alpha _{2}  \right) \rho^{3}+\cdots \right]. \hspace*{7mm}
\eea

\section{Reissner-N\"{o}rdstrom Black Hole}
\label{Reissner}

For Reissner-Nordstr\"{o}m black hole (static electrical charge $Q$), \cite{Reissner,Akad},
\begin{equation}
d s^2=\left[1-\frac{2M}{r}+\frac{Q^2}{r^2}\right] d t^2 -\frac{d r^2}{\left[1-\frac{2 M}{r}+\frac{Q^2}{r^2}\right]}- r^2  d \Omega ^{2}.
\end{equation}
This metric has two possible outer and inner horizons \cite{Reissner,Akad}, $r=M \pm \sqrt{M^2-Q^2}$.
The event horizon shrinks with increasing charges but another black hole likely appears near the singularity. With increasing charges, the two horizons become closer and the inner event horizons get larger, while the outer horizons shrink \cite{Kourosh}. With positive $ \sqrt{M^2-Q^2}$, the corresponding radius is determined by the outer horizon. At $M=r_s/2$,
\begin{equation}
\label{rs}
r=\frac{r_s}{2} \pm \sqrt{\frac{{r_s}^2}{4}-Q^2} \approx r_s\left[ 1 -\frac{Q^2}{r_s^2}-\frac{Q^4}{{r_s}^4}\right],
\end{equation}
where $r_s$ is Schwarzschild radius. If $Q$ entirely vanishes, Schwarzschild black hole metric shall be obtained.   

According to Bekenstein argument, the uncertainty in energy $\Delta E$ is defined with respect to its position uncertainty $\Delta E \approx  1/\Delta x$ \cite{Scardigli:2002}. We suppose $\Delta x\sim r$ and therefore,  
$\Delta E \geq r_s\left[ 1 - Q^2/r_s^2 - Q^4/{r_s}^4\right]^{-1}$ \cite{Bekensteina,Bekensteinb,Bekensteinc}. 
A minimal increase in the apparent horizon area $\Delta A\geq 8 \pi \ell _p ^2 E\, R$ is likely, when the black hole absorbs a particle of  energy $E$ and size $R$,  $(\Delta A)_{min} \geq 8 \pi \ell _p ^2 \Delta E\delta x$ \cite{Amelino2006}. Thus, 
\begin{eqnarray}
\Delta  A &\geq & 8 \pi \ell _p ^2 \left[ 1 -\frac{Q^2}{r_s^2}-\frac{Q^4}{{r_s}^4}\right]^{-1 }, \label{Adelata} \\
\frac{dS}{d A} &\approx & \frac {\Delta S_{min}}{\Delta
A_{min}}\simeq \frac{1}{4}\left[1-\frac{Q^2}
{{r_s}^2}-\frac{Q^4}{{r_s}^4}\right], \label{entrop}
\end{eqnarray}
and therefore, the Reissner-N\"{o}rdstrom  horizon area  can be expressed as function of the Schwarzschild one  \cite{Kourosh}
\begin{equation}
A=A_s - 8 \pi Q^2-\frac{(4 \pi)^2 Q^4}{A_s}+\cdots,
\end{equation}
where $A_s$ is the event horizon area corresponding to the Schwarzschild radius $r_s$. Then, the entropy  reads
\begin{eqnarray}
S \simeq \frac{A_s}{4}-\pi Q^2 \ln{\frac{ A_s}{4}}+ \frac{(\pi Q^{2})^{2}}{3}\Big(\frac{4}{A_{s}}\Big)+\cdots 
  \simeq S_{s}- (\pi Q^2)\, \ln{S_{s}}+ \frac{(\pi Q^{2})^{2}}{3} \, \Big(\frac{1}{S_{s}}\Big)+ \cdots. \label{areastand}
\end{eqnarray}

From Eq.  (\ref{rhoi}), the energy density can computed 
\bea
\rho &=& \rho _{s} -\frac{4}{3}\, (\pi Q^{2}) \, \rho ^{2}_{s} -\cdots,
\eea
and thus Hawking temperature  
\bea
T & \simeq & \frac{M}{2\pi \left[r_s
-\frac{Q^2}{r_s}-\frac{Q^4}{{r_s}^3} \right]^2} \simeq \frac{1}{8\pi M} \Big(1+ \frac{Q^2}{2
M^2}+\frac{5}{16}\frac{Q^4}{M^4}\Big).
\eea
With $T_{s}=1/(8 \pi M)$, the previous expression can be rewritten as
\bea 
T \simeq T_{s}\, +32\,(\pi Q)^{2}\, T^{3}_{s} +1280\,(\pi ^{2} Q)^{2}\,T^{5}_{s}.
\eea
Accordingly, the specific heat capacity is given as
\bea
C_{v}&=&12 \pi  T-192 \pi ^3 Q^2\,-1024 \pi ^5 Q^4 T^5 T^3-245760 \pi ^7 Q^6 T^7 =
c_{v_{s}} -\frac{Q^2}{27} \left\lbrace c_{v_{s}}^3 +\frac{1}{9} Q^2 c_{v_{s}} ^5 +\frac{5}{27} Q^4 c_{v_{s}}^7\right\rbrace,
\eea
where $ c_{v_{s}}$ is Schwarzschild specific heat capacity.

\subsection{Quadratic GUP and Black Hole Thermodynamics}

Because of $\Delta x \approx r$, the horizon area and electric change read $A =4\pi r_{+}^{2}$ and $Q\ll r$, respectively.
From the quadratic GUP approach, the modified horizon area of charged Reissner-N\"{o}rdstrom black hole can be given as function of the standard horizon area  
\bea
A_{GUP}=\, A+ \alpha ^{2} \pi \ln{A} -2 \alpha ^{2}  \pi  \left(\frac{Q^{2}}{2A}+\frac{Q^{4}}{A^{2}}\right)-32(\alpha ^{2} \pi)^{2} \left(\frac{1}{16A} + \frac{Q^{2}}{A^{2}} +\frac{2}{3}\frac{Q^{4}}{A^{3}}\right)+\cdots.
\eea
By using Bekenstein-Hawking entropy, the modified entropy and energy density, respectively,  read
\bea 
S_{GUP}&=& s+ \frac{\alpha ^{2}\pi}{4} \ln{s} -\, \alpha ^{2} \pi ^{2} \left(\frac{Q^{2}}{s} +2\frac{Q^{4}}{s^{2}}\right)-\,(\alpha ^{2} \pi)^{2}\left(\frac{1}{8s}+\frac{Q^{2}}{2s^{2}}+\frac{Q^{4}}{12s^{3}}\right)+\cdots+C, \label{Qentropy} \\
\rho _{GUP} &=& \frac{3}{2A} +\, \frac{3\, \alpha ^{2} \pi}{4A^{2}} + \frac{3 (\alpha ^{2} \pi )}{4} \left[ \frac{1}{A^2} + \frac{32 Q^2}{3 A^3} +\frac{128 Q^2}{A^4}  \right]+ 8 (\alpha ^{2} \pi)^2 \left[ \frac{1}{A^3} + 48 Q^2\frac{1}{A^4}+12 \frac{Q^4}{A^5} \right]+ \cdots,
\eea
where $C$ is an arbitrary constant.

From the standard entropy, Eq.   (\ref{3}), we get the modified energy density in terms of the standard energy density  
\bea 
\label{energys}
\rho_{GUP}=\rho +\, \frac{ \alpha ^{2} \pi}{3} \rho ^{2} +\frac{64( \alpha ^{2} \pi \,)}{27}\left(\alpha ^{2} \pi  +Q^{2}\right) \rho ^{3}+\frac{512}{27} {( \alpha ^{2} \pi ) Q^{2}}\left(2( \alpha^{2} \pi)+Q^{2} \right) \rho ^{4}+\frac{1024}{81} ( \alpha ^{2} \pi)^{2} {Q^{4}} \rho^{5}+C.
\eea

Again for charged Reissner-N\"{o}rdstrom black hole, Hawking temperature can be deduced from first law of thermodynamics, $T\, ds=dM$, 
\bea
T_{GUP} &=& \frac{1}{8 \pi M}\left\lbrace 1+\frac{(\alpha  \pi)^2}{ 16 \pi ^2 M^{2}}+\, \frac{(\alpha \pi)^{2}}{2 \pi} \left(\frac{Q^{2}}{8 \pi M^{4}}+\frac{Q^{4}}{8 \pi ^{2} M^{6}} \right)+(\alpha \pi)^{4}\, \left(\frac{1}{128 \pi ^{2}\, M^{4}}+\frac{Q^{2}}{64 \pi^{3} M^{6}}+\frac{Q^{4}}{1024 \pi ^{4} M^{8}}\right)\right\rbrace ^{-1}, \\ 
T_{GUP} &=& T \left\lbrace 1-4 (\alpha  \pi )^{2} \, T^{2} -32\left\lbrace (\alpha \pi)^{2}\left(1+8\pi (\alpha  \pi )^{2} Q^{2}\right) T^{4}+\, 128 (\alpha  \pi )^{2} \pi ^{3} Q^{2}   \left(1+ 4 \pi(\alpha  \pi )^{2} Q^{2}\right)\, T^{6}+ \, 512 \pi ^{4}(\alpha \, \pi )^{4}\, T^{8}\right\rbrace \right\rbrace  . \hspace*{6mm}
\eea

Relative to energy density, Eq.   (\ref{energys}), the geometric correction of the specific heat capacity based on quadratic GUP approach reads  
\bea
C_{v_{GUP}} &=& 12 \pi T + 48  (\alpha \pi)^2 (\pi T^3 +64 Q^2 \pi ^2 T^5 + \cdots) +3072 (\alpha \pi )^4 ( \pi T^5 +128 \pi T^7 +\cdots)+ \cdots,\\
C_{v_{GUP}} &=&C_v  +\frac{1}{36} \alpha ^2 C_v^3  +\frac{\alpha ^2 \left(\pi  \alpha ^2+Q^2\right)}{81 \pi} \, C_v^5 \,+\frac{4 \alpha ^2 Q^2  \left(2 \pi  \alpha ^2+Q^2\right)}{729 \pi ^2}\,C_v^7\,+\cdots.
\eea

In FLRW Universe, the modified Friedmann second equation due to interchange of quantum entropy, Eq.   (\ref{fred}), can be obtained from the derivative of Eq.   (\ref{Qentropy}) with respect to $A$; the horizon area. By using  Bekenstein-Hawking entropy, we get 
 \bea
\left(\dot{H} - \frac{K}{a^{2}} \right)\left[1-\frac{\alpha ^2 \pi}{4s} +(\alpha \pi )^{2} \left[ \frac{Q^{2}}{s^{2}}+4\frac{Q^{4}}{s^{3}}\right]+(\alpha \pi )^{4} \left[\frac{1}{8 s^{2}}+ \frac{Q^{2}}{s^{3}}+\frac{Q^{4}}{s^{4}}\right]+ \cdots\right]=\,-16 \pi \nn \, \\{} \left[p  +\rho + \frac{64}{27} (\alpha \pi )^2  Q^2 \left[\rho ^{3} + 8 Q^2 \rho ^{4}  \right]+ \frac{512}{27}(\alpha \pi )^4 \left[\frac{1}{8}\rho ^{3} + 2 Q^2 \rho ^{4}+ \frac{2 Q^4}{3} \rho ^5  \right]+ \cdots \right]. & & \hspace*{8mm}\label{RessinerGUP} 
\eea

\subsection{Linear GUP and Black Hole Thermodynamics}

In $\mathcal{O(\alpha)}$, the uncertainty of momentum, Eq. (\ref{LGup}), implies geometric correction to the quantum thermodynamics of the charged Reissner-N\"{o}rdstrom black hole
\bea
dA_{GUP} &=& \frac{(A-4\pi Q^{2})dA}{A-4\pi Q^{2}+2\, \alpha \sqrt{\pi}\sqrt{A}}, \\
A_{DSR-GUP}&=&\, A -4 \alpha \sqrt{\pi}\sqrt{A}+\, 4\alpha ^{2}\pi \ln{(\sqrt{A}(2\alpha \sqrt{\pi}+\sqrt{A}})-4 \pi Q^{2})-\nn \, \\ 
&&  \frac{4\pi ^{3/2} \alpha (\alpha ^{2}+2\pi Q^{2})}{\sqrt{\pi \alpha ^{2}+4\pi Q^{2}}} \ln{\left(\frac{-\sqrt{A} +2\alpha \sqrt{\pi }-2\sqrt{\pi \alpha ^{2}+4\pi Q^{2}}}{-\sqrt{A} +2\alpha \sqrt{\pi }+2\sqrt{\pi \alpha ^{2}+4\pi Q^{2}}}\right)}.
\eea

For finite $\alpha$ and $Q\ll\sqrt{A}$, the entropy reads 
\bea 
\label{slinear}
S_{DSR-GUP}=s-\,2 \alpha \sqrt{\pi}\sqrt{s}+\, \alpha ^{2}\pi \ln{({s}-\pi Q^{2})}-\, \alpha \pi \frac{(\alpha ^{2}+2\,Q^{2})}{\sqrt{\alpha ^{2}+4Q^{2}}} \ln{\left(\frac{-\sqrt{s}+\alpha \sqrt{\pi}-\sqrt{\pi \alpha ^{2}+4\pi Q^{2}}}{-\sqrt{s}+\alpha \sqrt{\pi}+\sqrt{\pi \alpha ^{2}+4\pi Q^{2}}}\right)}+C,
\eea
where $C$ is an arbitrary constant.

From Eq.   (\ref{rhoi}) and the derivative of Eq.   (\ref{slinear}) with respect to the unmodified entropy, the energy density reads
\bea
\rho_{DSR-GUP}&=&\left(1- \frac{\alpha ^{2}}{\pi Q^{2}}\right)\rho\,-\frac{4\sqrt{2}}{3\sqrt{3}} \alpha\, \pi \, \rho ^{3/2}+ \frac{3 \alpha ^{2}}{8 \pi ^{3}Q^{4}} \ln{\left(\frac{3}{2 \,\rho}\right)}+ \nn \\
& & \eta \, \left[3\, \ln{\left(\frac{3}{2 \rho}\right)}+\left(+ \alpha \, \sqrt{\pi} +\, \sqrt{{\alpha ^{2}\pi + 4\, \pi Q^{2}}}\right)\ln{\left((\frac{3}{2 \rho})^{2}-16 (\alpha \sqrt{\pi} + \sqrt{{\alpha ^{2}\pi\,+\,4 \pi Q^{2}}})\right)}\right]+
\nn \, \\ 
&& \eta \, \left[\left(-\, \alpha \, \sqrt{\pi} + \sqrt{{\alpha ^{2}\pi+4 \pi Q^{2}}}\right)\ln{\left((\frac{3}{2 \rho})^{2}-16 (-\, \alpha \sqrt{\pi} + \sqrt{{\alpha ^{2}\pi+4 \pi Q^{2}}})\right)}\right], \label{wwww}
\eea
where $\eta =3 \alpha \pi (\alpha ^{2}+2Q^{2})\sqrt{\alpha ^{2}\pi +4 \pi Q^{2}}/4[(\alpha \pi)^{2}-\, (\alpha ^{2}\pi +4 \pi Q^{2})] \sqrt{\alpha ^{2}+4Q^{2}}$.

The Hawking temperature frome linear approach of the GUP can be computed as
\bea
T_{DSR-GUP} &=& \frac{1}{8\pi M}\left[1-\frac{\alpha}{2}-\frac{\alpha ^{2}}{4M^{2}-Q^{2}}-\frac{\alpha(\alpha ^{2}+\,2Q^{2})}{8M(Q^{2}+M(\alpha -\,M))}\right]^{-1}, \\ 
T_{DSR-GUP} &=& \left(1+\frac{\alpha}{2}\right)\,T+16\alpha \pi \left[ \left(\alpha \, \pi +(\alpha ^{2}+2\,Q^{2})\right)T^{3}+ \left(16\, \alpha \, \pi Q^{2} +\frac{Q^{2}}{\alpha T-\,8 \,\pi}\right)T^{5}\right].
\eea

The geometric correction of specific heat capacity because of corrected Hawking radiation temperature from linear GUP  reads 
\bea
C_{v_{DSR-GUP}}&=& -48 \pi ^{5/2} \alpha  T^2 \,-\frac{3 \alpha ^2}{4 \pi ^3 Q^4 T}+12 \pi  T \left(1-\frac{\alpha ^2}{\pi 
   Q^2}\right)- \eta \frac{6}{T}- \nn \\
   && \eta \left(\frac{\sqrt{\pi  \alpha ^2+4 \pi 
   Q^2}-\sqrt{\pi } \alpha }{4 \pi ^2 T^5 \left(\frac{1}{16 \pi ^2 T^4}-16 \left(\sqrt{\pi 
   \alpha ^2+4 \pi  Q^2}-\sqrt{\pi } \alpha \right)\right)} + \right. \nn \\
   && \left. \hspace*{5mm}
   \frac{\sqrt{\pi } \alpha
   +\sqrt{\pi  \alpha ^2+4 \pi  Q^2}}{4 \pi ^2 T^5 \left(\frac{1}{16 \pi ^2 T^4}-16
   \left(\sqrt{\pi } \alpha +\sqrt{\pi  \alpha ^2+4 \pi 
   Q^2}\right)\right)}\right).
  \eea
The geometric correction of specific heat capacity is
  \bea
C_{v_{DSR-GUP}} &=& c_v \left(1-\frac{\alpha ^2}{\pi  Q^2}\right)-\frac{1}{3} \sqrt{\pi } \alpha  c_v^2\,-\frac{9 \alpha ^2}{\pi ^2 Q^4 c_v}+ \nn \\
&& \eta  \left(-\frac{62208 \pi ^3 \left(\sqrt{\pi  \alpha
   ^2+4 \pi  Q^2}-\sqrt{\pi } \alpha \right)}{c_v^5 \left(\frac{1296 \pi ^2}{c_v^4}-16
   \left(\sqrt{\pi  \alpha ^2+4 \pi  Q^2}-\sqrt{\pi } \alpha \right)\right)}- \right. \nn \\
   && \left. \hspace*{8mm} \frac{62208 \pi
   ^3 \left(\sqrt{\pi } \alpha +\sqrt{\pi  \alpha ^2+4 \pi  Q^2}\right)}{c_v^5
   \left(\frac{1296 \pi ^2}{c_v^4}-16 \left(\sqrt{\pi } \alpha +\sqrt{\pi  \alpha ^2+4 \pi 
   Q^2}\right)\right)}-\frac{72 \pi }{c_v}\right).
   \eea

The modified Friedmann second equation is given as 
\bea 
\left(\dot{H} - \frac{K}{a^{2}} \right) \left[1-2\frac{\alpha \pi}{\sqrt{s}}+\frac{\alpha ^{2}\pi}{s-\pi Q^{2}} - 2 \alpha \pi  \frac{\sqrt{\alpha ^{2}\pi +4\pi Q^{2}}\, \sqrt{\alpha ^{2}+2Q^{2}} }{ 4 \pi Q^{2} \sqrt{s}} +\alpha \sqrt{\pi}s -2s^{3/2}+ \cdots\right]=-16 \pi  \nn \, \\{}  \left[p+ \rho -\frac{4\sqrt{2}}{3\sqrt{3}} \alpha\, \pi \, \rho ^{3/2}- \frac{\alpha ^{2}}{\pi Q^{2}}\rho\,+ \frac{3 \alpha ^{2}}{8 \pi ^{3}Q^{4}} \ln{\left(\frac{3}{2 \,\rho}\right)}+ \cdots \right].\label{RessinerDSR}
\eea

\subsection{MDRs and Black Hole Thermodynamics}

From Eq.   (\ref{energymdr}), where $\Delta x =\sqrt{\frac{A}{\pi}}\left(1-\frac{4 \pi Q}{A}\right)$ and by using Eq.   (\ref{smdr}),  the geometric correction of of charged Reissner-N\"{o}rdstrom entropy
\bea
dS_{MDR} &=& \frac{1}{4}\left[1+\, \frac{3 \alpha _{1}}{2} \frac{1}{(\Delta x)^{2}}- \frac{5}{8}\, \left( \alpha _{1}^{2} - 4\alpha _{2} \right)\frac{1}{(\Delta x)^{4}}\right]d\,A, \\
S_{MDR} &=& \frac{1}{4}\left[A+\, \frac{3}{2} \alpha _{1} \pi (\,1 + 8\, \pi Q)\ln{(A)}+5 \pi ^{2}\, (\alpha _{1}^{2} -\,4 \alpha _{2})\left(\frac{1}{8A}+\,\frac{\pi Q}{2\,A^{2}}\,-\frac{2 \pi \,Q^{2}}{A^{3}}\right)+\frac{72 \alpha _{1} (\pi Q)^{2}}{A}+\cdots+C\right], \hspace*{6mm}\label{sAMDE2}
\eea
where $C$ is an arbitrary constant.
By using the Bekenstein-Hawking entropy-area law \cite{Bekensteina,Bekensteinb,Bekensteinc,Hawking}, the modified entropy reads
\bea 
S_{MDR}=s+\frac{3}{8} \alpha _{1} \pi (\,1 + 8\, \pi Q)\ln{(s)}+\frac{9}{2} \alpha _{1}\pi \frac{\pi  Q^{2}}{s}+\frac{5 \pi ^{2}}{4}\, (\alpha _{1}^{2} -\,4 \alpha _{2})\left(\frac{1}{32s}+\frac{\pi Q}{32 s^{2}}-\frac{\pi Q^{2}}{32 s^{3}}+\cdots \right).
\eea

Again from the Bekenstein-Hawking entropy-area law \cite{Bekensteina,Bekensteinb,Bekensteinc,Hawking} and Eq. (\ref{sAMDE2}), we obtain
\bea 
A_{MDR}=\left[A+\, \frac{3}{2} \alpha _{1} \pi (\,1 + 8\, \pi Q)\ln{(A)}+\, \frac{72 \alpha _{1} (\pi Q)^{2}}{A}+5 \pi ^{2}\, (\alpha _{1}^{2} -\,4 \alpha _{2})\left(\frac{1}{8A}+\,\frac{\pi Q}{2\,A^{2}}\,-\frac{2 \pi \,Q^{2}}{A^{3}}\right)+\cdots\right]. \label{AAMDE3}
\eea

The differentiate Eq.   (\ref{sAMDE2}) and the integration of Eq.   (\ref{rhoi}) lead to 

\bea
\left(\dot{H} - \frac{K}{a^{2}} \right)\left[1+\frac{3 \alpha _{1} \pi}{2} \left[ \frac{1+8 \pi Q}{4 s} -\, \frac{3 \pi Q^{2}}{s^{2}}\right]-\, \frac{5 \pi ^{2}}{128} \left(\alpha _{1}^{2}-4 \alpha _{2}\right) \left[\frac{1}{ s^{2}}+\frac{2 \pi Q}{s^{3}}-\frac{3 \pi ^{2} Q^{2}}{s^{4}}\right] + \cdots \right] = & &\nn \\
-16 \pi \left[ p+\rho +\frac{\alpha _{1} \pi}{2} \left[(1+24 \pi Q)\rho ^{2}-\, \pi Q^{2} \rho ^{3}\right]+\frac{32}{27}\pi ^{2} (\alpha _{1}^{2} -\,4 \alpha _{2})\left[\frac{5}{2}\rho ^{3}+\,\frac{5 \pi Q}{16} \rho ^{4} -\,\pi Q^{2} \rho ^{5} \right]+\cdots \right]. & & \hspace*{8mm} \label{MDRRHO1}
\eea

Taken into account Eq.  (\ref{4}), Hawking temperature is given as 
\bea
T_{MDR}&=& T\left[1-\frac{3}{32} \alpha _{1}\left(\frac{(1+ \pi Q)}{M^{2}}-\frac{3\,Q^{2}}{M^{4}}\right)-\,\frac{5}{32}\left(\alpha _{1}^{2}-4 \alpha _{2}\right) \left( \frac{1}{64 M^{4}}+\frac{Q}{128 M^{6}}-\frac{3 Q^{2}}{1024\pi M^{8}}\right)\right]+C, \\
T_{MDR}&=& T\left[1-{3} \alpha _{1} \pi ^{2}\left(2{(1+ \pi Q)}\,T^{2}-384 \pi ^{2}\,Q^{2}\,T^{4}\right)+\,{5}\pi ^{4} \left(\alpha _{1}^{2}-4 \alpha _{2}\right) \left( 2\,T^{4}+64 \pi ^{2}{Q}\,T^{6}-\,512 \pi ^{4} Q^{2} T^{8}\right)\right]. \hspace*{7mm}
\eea

By using the energy density as a label of horizon area (\ref{MDRRHO1}), the specific heat capacity leads to 
\bea
C_{v_{MDR}} &=& 12 \pi T+\, 24 \alpha _{1}\pi ^{3}\left[(1+24 \pi Q) T^{3}\, -27 \pi ^{2}Q^{2} T^{5}-640\pi ^{5}  (\alpha _{1}^{2} -\,4 \alpha _{2})\left(6\, T^{5}+ 6 \pi ^{2} Q T^{7}-\, \ldots \mathcal{O}(T^{9})\right)\right], \hspace*{7mm}\\
C_{v_{MDR}} &=& c_{v}+\alpha _{1} \left( \frac{1+24 \pi Q}{72} c_{v}^{3}-\, \frac{Q^{2}}{384} c_{v}^{5}\right)-\frac{5}{324}(\alpha _{1}^{2} -\,4 \alpha _{2})\left(c_{v}^{5}\,+\frac{Q}{144}c_{v}^{7}-\ldots\mathcal{O}(c_{v}^{9})\right).
\eea

Finally, the modified Friedmann second equation reads  
\bea
\left(\dot{H} - \frac{K}{a^{2}} \right)\left[1+\frac{3 \alpha _{1} \pi}{2} \left[ \frac{1+8 \pi Q}{4 s} -\, \frac{3 \pi Q^{2}}{s^{2}}\right]-\, \frac{5 \pi ^{2}}{128} \left(\alpha _{1}^{2}-4 \alpha _{2}\right) \left[\frac{1}{ s^{2}}+\frac{2 \pi Q}{s^{3}}-\frac{3 \pi ^{2} Q^{2}}{s^{4}}\right] + \cdots \right] = & &\nn \\
-16 \pi \left[ p+\rho +\frac{\alpha _{1} \pi}{2} \left[(1+24 \pi Q)\rho ^{2}-\, \pi Q^{2} \rho ^{3}\right]+\frac{32}{27}\pi ^{2} (\alpha _{1}^{2} -\,4 \alpha _{2})\left[\frac{5}{2}\rho ^{3}+\,\frac{5 \pi Q}{16} \rho ^{4} -\,\pi Q^{2} \rho ^{5} \right]+\cdots \right]. & & \hspace*{8mm} \label{RessinerMDR}
\eea


\section{Garfinkle-Horowitz-Strominger Black Hole}
\label{Garfinkle}
The metric of  Garfinkle-Horowitz-Strominger dilated black hole is
\bea
ds^{2}=\left(1-2\frac{M}{r}\right)\, dt^{2}+\, \left(1-2\frac{M}{r}\right)^{-1} dr^{2}\,+r(r-2a)\, d\Omega ^{2},
\eea
where $a=Q^{2}e^{-2 \phi} _{0}/(2M)$, is the asymptotic value of the dilation field, $M$ is the mass, and $Q$ is magnetic charge of the balck hole. When $a$ is constant, the horizon area $A$, entropy $S$, energy density $\rho$, Hawking radiation temperature $T$, and Specific heat $C_v$ respectively, reads
\bea
A &=& 4\pi r_{H}(r_{H}-2a)=\,16 \pi M\,(M-a), \label{11} \\
S &=& \pi  r_{H}(r_{H}-2a)=\,4 \pi M\,(M-a)=\frac{A}{4}, \label{22} \\
\rho &=& \frac{3}{8 \pi r_{H} (r_{H}-2a)}=\, \frac{3}{32 \pi M (M-\,a)}=\frac{3}{2\,A}, \label{33} \\
T &=& \frac{1}{4\pi (r_{H}-2a)} =\, \frac{1}{4\pi (2M-a)}, \label{44} \\
c_v&=&12 \pi \, T=\, \frac{3}{ (r_{H}-2a)} =\, \frac{3}{(2M-a)}. \label{55} 
\eea
Again, $\Delta x=2\,r_{H}$ and $a \ll  r_H$. The quantum geometric corrections Garfinkle-Horowitz-Strominger black hole  thermodynamics shall be estimated in following sections.

\subsection{Quadratic GUP and Black Hole Thermodynamics}

If $a\rightarrow 0$, the horizon area of uncharged de Sitter-Schwarzschild black hole is given as
\bea
A_{GUP}=A+\, (\alpha ^{2}\pi ) \left[ \ln{(A)}+8\left(\sqrt{\frac{a^{2}\pi}{A}}\right)+8\left({\frac{a^{2}\pi}{A}}\right)\right]-2(\alpha ^{2}\pi)^{2}\left[\frac{1}{A}-\frac{8\,a^{2} \pi}{A^{2}}\right]+\cdots .
\eea
From the entropy-area relation, the quantum geometric corrections to the entropy, Eq.   (\ref{22}), 
\bea
S_{GUP}=\, s+\, (\alpha ^{2}\pi) \left[ \ln{(s)}+\left(\sqrt{\frac{a^{2}\pi}{s}}\right)+\left({\frac{a^{2}\pi}{s}}\right)\right)-\frac{( \alpha ^{2}\pi)^{2}}{8}\left(\frac{1}{s}-\frac{2\,a^{2}\pi}{s^{2}}\right)+\cdots+C,
\eea
where $C$ is an arbitrary constant.

According to corrected entropy, a well definition for the corresponding quantum geometrical correction to the energy density relation, Eq.    (\ref{33}), reads
\bea
\rho _{GUP}= \rho +{(\alpha ^{2}\pi)}\, \left( \frac{1}{3} \rho ^{2}\,\frac{32}{27} a^2 \rho ^3 - \frac{16\sqrt{6\pi}}{45}\,a\, \rho ^{5/2} \right) \,
-\frac{8}{27} ( \alpha ^{2}\pi )^{2} \left( \rho ^3 - 8 a^2 \rho ^4 \right)+\cdots, \label{rhoGran}
\eea
From Eq.   (\ref{22}) and the inequality $a \ll  r_H$, the Hawking temperature turns to be subject of modification
\bea
\frac{dS_{GUP}}{dM} &=& 8\pi M +\alpha ^{2} \pi \left( \frac{2}{M} -\,\frac{a}{2\,M^{2}}-\,\frac{a^{2}}{2\,M^{3}}\right)+\frac{(\alpha ^{2} \pi)^{2}}{16\pi} \left(\frac{1}{M^{3}}-\frac{a^{2}}{ M^{5}}\right), \\
T_{GUP} &=& \frac{1}{8 \pi M}\left[1-\frac{\alpha ^{2} \pi}{8\pi}   \left( \frac{2}{M^{2}} -\,\frac{a}{2\,M^{3}}-\,\frac{a^{2}}{2\,M^{4}}\right)-\, \frac{(\alpha ^{2} \pi)^{2}}{128 \pi ^{2}}\left( \frac{1}{M^{4}}-\frac{a^{2}}{ M^{6}}\right)\right], \\
T_{GUP} &=& T\, \left[1-16\, \alpha ^{2} \pi ^{2} \left(\,T^{2}-2\pi a T^{3}-16 \pi ^{2} a^{2} T^{4} \right)-32 ( \alpha ^{2} \pi ^{2})^{2} \left(T^{4}-64 \pi a^{2} T^{6}\right)\right].
\eea

From modified energy density of  Garfinkle-Horowitz-Strominger black hole, Eq. (\ref{rhoGran}), the specific heat capacity can be deduced
\bea
C_{v_{GUP}}&=& 12 \pi  T +48 \pi ^3 \alpha ^2 T^3\,-384 \pi ^4 a\, \alpha ^2 T^{4}\,- 384 \pi ^5 \, \left(\alpha ^{2}-4\,a^{2}\right) \alpha ^4 T^5+24576 (\alpha ^{2} \pi ^{3})^2 a^{2} \pi T^7, \\
C_{v_{GUP}} &=& C_{v}+ \frac{\alpha ^{2}}{36} \,C_{v}^{3} -\frac{1}{54} \alpha ^{2} a \, C_{v}^{4} -\, \frac{\alpha ^{4}}{648}\left(\alpha ^{2}-4\,a^{2}\right) C_{v}^{5}+\cdots.
\eea

According to the quantum geometric correction to entropy and by applying Eq.  (\ref{fred}), the modified Friedmann second equation reads
\bea
\left(\dot{H} - \frac{K}{a^{2}} \right)\left[1+ \frac{\alpha ^2 \pi}{2}  \left[\frac{1}{2s}-\frac{\sqrt{\pi } a}{s^{3/2}} -\frac{\pi  a^2}{ s^2}\right]+(\frac{\alpha ^2 \pi}{2})^2 \left[\frac{1}{4 s^2}-\frac{\pi  a^2}{ s^3}\right] +\cdots \right] = & & \nn \\ 
-16 \pi \left[p+  \rho + \alpha ^{2}\pi \left[ \frac{1}{3} \rho^2  - \frac{16\sqrt{6 \pi\, a}}{45} \rho ^{5/2} \right] - \frac{64}{27} (\alpha ^{2}\pi)^2 \left[ \frac{(\alpha ^2 - 4 a^2)}{4} \rho ^{3}  - a^2	\pi \rho ^4 \right] + \cdots\right]. \label{GarfinkleGUP}
\eea

\subsection{Linear GUP and Black Hole Thermodynamics}

In $\mathcal{O(\alpha)}$, the uncertainty in momentum, Eq. (\ref{LGup}), implies  quantum geometrical corrections to the black hole thermodynamics. The change in the area of black hole is given as
\bea
dA_{DSR-GUP}=\left(1+\frac{2\alpha}{\Delta x}\right)^{-1}\,dA.
\eea
The horizon area of the charged black hole can be calculated as 
\bea
A_{DSR-GUP}= A -4\, \alpha \sqrt{\pi} \sqrt{A}+8\, \alpha \pi(a+\alpha ) \, \ln{\left(\sqrt{\frac{A}{\pi}}+2(a+ \alpha)\right)}.
\eea

At $a\ll r_{H}$, the modified entropy from linear GUP approach is 
\bea
S_{DSR-GUP}= S\, -2\, \alpha \sqrt{\pi} \sqrt{s}+\, \alpha \pi(a+\alpha ) \, \ln{s}+C, \label{SDSRMDR3}
\eea
The energy density can be obtained by the differentiation of Eq.   (\ref{SDSRMDR3}), the integration of Eq.   (\ref{rhoi}) and taking into account Eq.  (\ref{33}), 
\bea
\rho_{DSR-GUP}= \rho - \frac{4\sqrt{2}}{3\sqrt{3}} \alpha \sqrt{\pi} \rho ^{3/2}+\frac{4}{3}\alpha \pi (a+\alpha)\rho ^{2}+C,
\eea
where $C$ is an arbitrary constant.

Because of the geometric correction of entropy, and by applying the first law of thermodynamic of black holes, we have 
\bea
T_{DSR-GUP} &=& \frac{1}{8 \pi M}\left(1+\frac{\alpha}{2 M} -\, \frac{\alpha (a+\alpha)}{4 M^{2}}\right), \\
T_{DSR-GUP} &=& T\left(1+ 4\alpha \pi T -16\alpha \pi ^{2}(a+\alpha ) T^{2} \right).
\eea
The calculation of the specific heat capacity is straightforward 
\bea
C_{v_{DSR-GUP}} &=& 12 \pi T - 48 \pi ^2  T \sqrt{T^2} \alpha +  16 \pi ^2 T \alpha (a +\alpha), \\
C_{v_{DSR-GUP}} &=& C_{v}\,+\frac{4}{3}\alpha \pi (\alpha +a) C_{v} -\frac{\alpha}{3} C_{v}^{2}.
\eea

For FLRW Universe, the modified second Friedmann equation reads 
\bea 
\left(\dot{H} - \frac{K}{a^{2}} \right)\left(1-\alpha \sqrt{\pi} \frac{1}{\sqrt{s}} +\alpha \pi (a+\alpha )\frac{1}{s}\right)&=&-16\pi \, \left(p+\left[ \rho - \frac{4\sqrt{2}}{3\sqrt{3}} \alpha \sqrt{\pi} \rho ^{3/2}+\frac{4}{3}\alpha \pi (a+\alpha)\rho ^{2}+C\right]\right). 
\eea

\subsection{MDRs and Black Hole Thermodynamics}

The quantum geometrical correction of charged black hole, where $\Delta x= 2 r_{H}$ and $r_{H}=\frac{1}{2}\, \sqrt{A/\pi}+\,a$, and using Eq.   (\ref{MDRALL}) can be obtained
\bea
A_{MDR}=A-\, \frac{3 \alpha _{1} \pi}{2} \ln{A}\, -\frac{12\,a\, \alpha _{1} \pi ^{3/2}}{\sqrt{A}} +\frac{5 \pi ^{2}}{2}\left(\alpha _{1}^{2} -\, 4 \alpha _{2}\right)\left(\frac{1}{4A}\,-\frac{4\, a \pi ^{1/2}}{A^{3/2}}\right).
\eea
From the entropy-area relation,  
\bea
\label{S0}
S_{MDR}=s-\, \frac{3 \alpha _{1} \pi}{8} \ln{s}\, -\frac{3\,a\, \alpha _{1} \pi ^{3/2}}{\sqrt{s}} +\frac{5 \pi ^{2}}{8}\left(\alpha _{1}^{2} -\, 4 \alpha _{2}\right)\left(\frac{1}{4s}\,-\frac{4\, a \pi ^{1/2}}{s^{3/2}}\right)+C,
\eea
The corrected  energy density reads
\bea
\label{S1}
\rho _{MDR}= \rho -\frac{\alpha _{1} \pi}{2} \rho ^{2} +\frac{5\sqrt{2}}{8\sqrt{3}} +\,{\alpha _{1} \pi ^{3/2} a} \rho ^{5/2} -\,\frac{5 \pi^{2}}{3} \left(\alpha _{1}^{2} -\, 4 \alpha _{2}\right) \left( \frac{\rho ^{3}}{18}-\frac{8\sqrt{\frac{2}{3}}}{21} a \pi ^{1/2} \rho ^{7/2}\right)+C,
\eea
Where $C$ is an arbitrary constant.

By applying the first law of thermodynamic for black holes $T\, dS=\, dM$,  the modification in Hawking radiation temperature reads
\bea
T_{MDR}=T\left[ 1+6 \alpha _{1} \pi ^{2} \left(1- \, 8 a T \right) T^{2}+10 \pi ^{2} \left(\alpha _{1}^{2} -\, 4 \alpha _{2}\right)  \left( \frac{\pi}{2}-64 a \pi ^{3} T \right) T^{3}\right].
\eea
Corresponding to the energy density, the specific heat capacity is modified as
\bea
C_{v{MDR}}= C_{v} -\frac{\alpha _{1}}{24}\, C_{v}^{3}+\frac{\alpha _{1} a}{36} C_{v}^{4}-\frac{5}{7776}\, \left(\alpha _{1}^{2} -\, 4 \alpha _{2}\right)  \, \left( \frac{4}{3}\, C_{v}^{5} -\, a C_{v}^{6} \right).
\eea

Finally, the modified Friedmann equation can be deduced from Eqs. (\ref{S0}), (\ref{S1}) and (\ref{fred}), 
\bea
\left(\dot{H} - \frac{K}{a^{2}} \right)  \left[1 -\frac{3 \pi \alpha _{1}}{8 s}+\frac{3 \pi ^{3/2} a \alpha _{1}}{4 s^{3/2}}-\frac{5
   \pi ^2 \left(\alpha _{1} ^2-4 \alpha _{2}\right)}{128 s^2} - \frac{5 \pi ^{5/2} a \left(\alpha _{1}^{2} -\, 4 \alpha _{2}\right)}{32 s^{5/2}}+\cdots \right] = & &\nn \\
   -16 \pi \left[p+ \rho -\frac{\alpha _{1} \pi}{2} \rho ^{2} +\frac{5\sqrt{2}}{8\sqrt{3}} +\,{\alpha _{1} \pi ^{3/2} a} \rho ^{5/2} -\,\frac{5 \pi^{2}}{3} \left(\alpha _{1}^{2} -\, 4 \alpha _{2}\right)   \left( \frac{\rho ^{3}}{18}-\frac{8}{21} \sqrt{\frac{2}{3}} a \pi ^{1/2} \rho ^{7/2}\right)+\cdots \right], && \label{GarfinkleMDR}
\eea

\section{Comparison between modified and non-modified thermodynamics}
\label{compare}

\begin{figure}[htb]
\centering{
\includegraphics[width=8cm,angle=0]{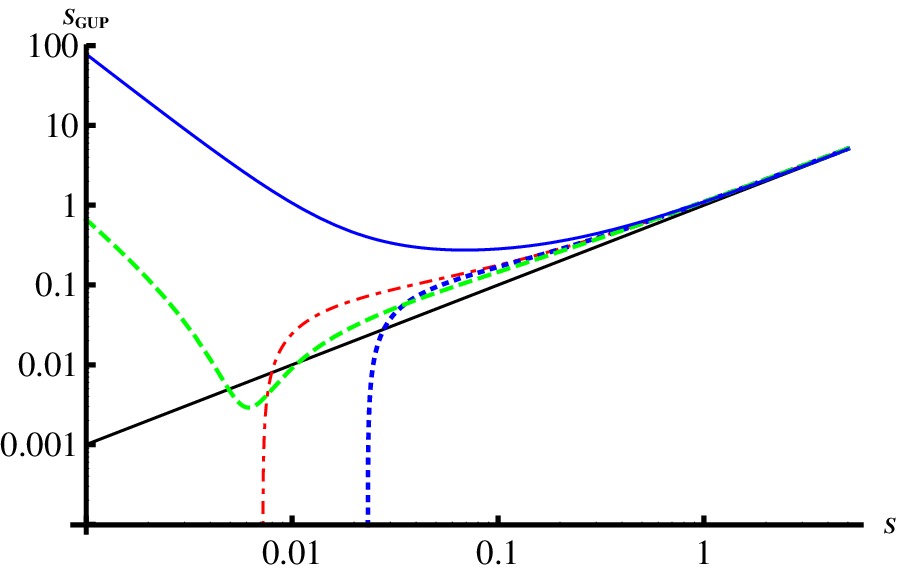}
\includegraphics[width=8cm,angle=0]{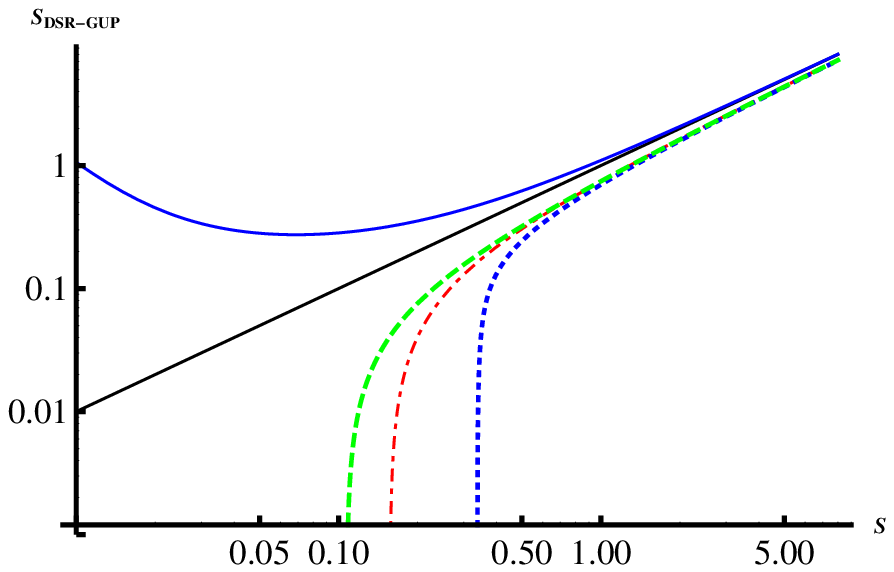}
\caption{Left-hand panel shows the modified entropy due to quadratic GUP  and MDR approaches in dependence on non-modified entropy. Right-hand panel gives the modification due to linear GUP approach. Dash-dotted, dotted and dashed curves give the results for de Sitter-Schwarzschild, Reissner-N\"{o}rdstrom and Garfinkle-Horowitz-Strominger at $\alpha=0.1$, respectively. Solid lines are there to compare with non-modified entropies. The solid curves represent standard de Sitter-Schwarzschild. 
\label{fig:1}}
}
\end{figure}

The modified entropy due to quadratic GUP and MDR approaches is given in dependence on non-modified entropy in left-hand panel of Fig. \ref{fig:1}. The modifications due to linear GUP approach is drawn in the right-hand panel. In all these calculations, the parameters $Q$, $a$, arbitrary constant $C$ and $\alpha =0.1$ are fixed. While GUP effect seems to disappear at relative small standard entropy, we find that the modified entropies of the different three-types of black holes from quadratic GUP and MDR reach the standard one obtained at large values of standard entropies. In other words, we can consider that the matter of each type of black holes loses its entropy over time. Apparently, this represents the second law of thermodynamics of the black holes. With the time, the black holes radiate causing a decrease in both mass and area of the horizon. 

The solid curved represent the standard de Sitter-Schwarzschild black hole. the solid lines are illustrated to compare with the non-modified entropy. The dash-dotted, dotted and dashed curves are the results from modified de Sitter-Schwarzschild, Reissner-N\"{o}rdstrom and Garfinkle-Horowitz-Strominger black holes, respectively. It is obvious that the modified entropies start below the reference line. Increasing standard entropy brings the modified entropies very close to the standard ones.  It is worthwhile to notice here that the modified entropy of Garfinkle-Horowitz-Strominger black hole starts with positive value where the other two types do not exist. In the right panel, the modified entropies of the three metric-types start to exist at finite standard entropy. 

\begin{figure}[htb]
\centering{
\includegraphics[width=8cm,angle=0]{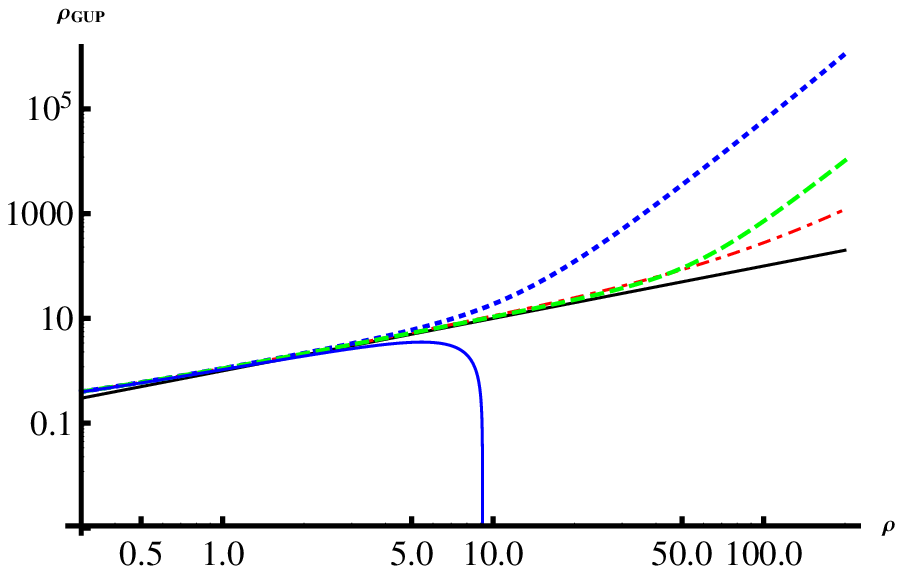}
\includegraphics[width=8cm,angle=0]{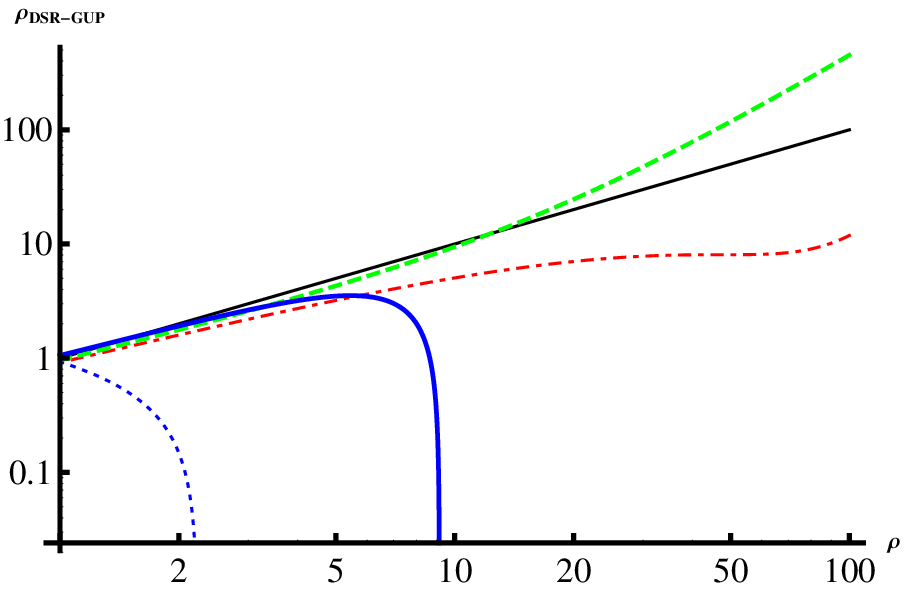}
\caption{The same as in Fig. \ref{fig:1} but for energy density. \label{fig:2} }
}
\end{figure}

In Fig. \ref{fig:2}, the modified energy densities of the three types of black holes are given in dependence on standard one. We notice that the modified energy densities are the same as the standard one, especially at low energy densities. From the first law of of thermodynamics of the black hole, the change of energy (proportional to mass) is related to change of the area, angular momentum and electric charge. The angular momentum and electric charge represent the change in energy due to rotation and electromagnetism properties of black holes, which apparently play an essential role in the stationary black holes to be compatible with the standard energy density at large values. In right panel, it is obvious that the effect of gravitational filed is able to overcome the electromagnetism properties of Reissner-N\"{o}rdstrom black hole delayed than its standard value.  It is obvious that the modified energy densities has the same value of the non-modified one at very low values. Increasing energy density separates modified entropies away from the standard value. The energy density of modified Garfinkle-Horowitz-Strominger black hole exceeds the standard value, while that of modified de Sitter-Schwarzschild and Reissner-N\"{o}rdstrom black holes remain below. In left-hand panel, this behaviour appears much faster than in the right-hand panel. As difference between the two approaches, we find that the modified energy density of Reissner-N\"{o}rdstrom black holes due to linear GUP and MDR remains finite (smaller than the standard line). It disappears in the other approach. Almost the same is found for Garfinkle-Horowitz-Strominger  black hole. The energy density of modified de Sitter-Schwarzschild remains finite in both approaches.

\begin{figure}[htb]
\centering{
\includegraphics[width=8cm,angle=0]{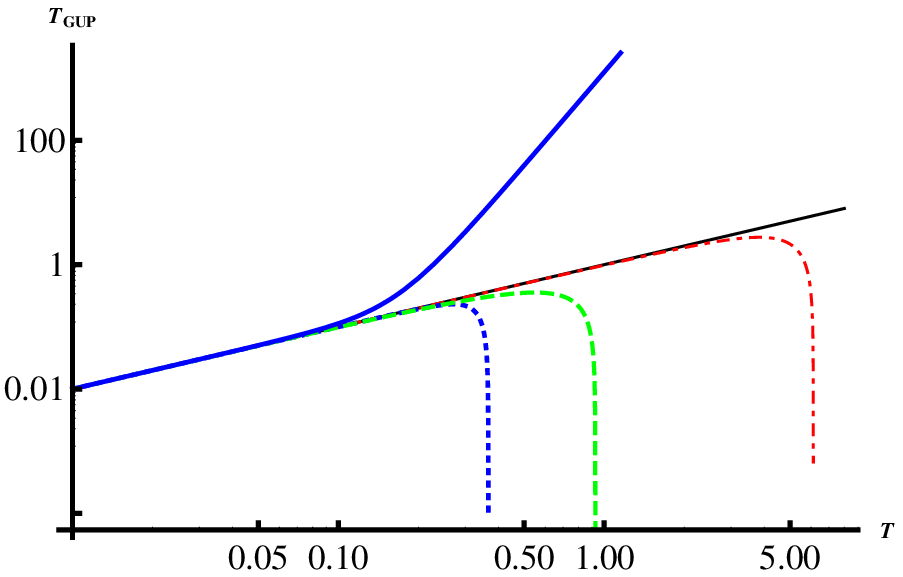}
\includegraphics[width=8cm,angle=0]{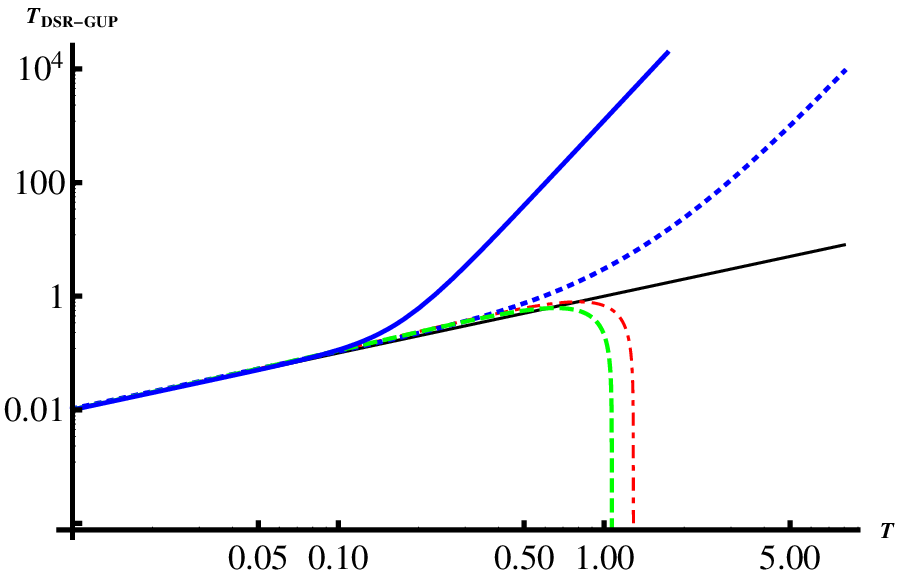}
\caption{The same as in Fig. \ref{fig:1} but for Hawking temperature. 
\label{fig:3} }
}
\end{figure}

In Fig. (\ref{fig:3}), Hawking temperature of modified black holes is given in dependence on the non-modified one. We find that the modified Hawking temperatures of the different types of black holes are the same as the standard one, especially at low values of the standard Hawking temperature. It is worthwhile to notice that Hawking radiation might have been existed in the primordial stage of Universe. At very high standard Hawking temperature (small-mass black holes), the modified temperature drops at different values which means that the Hawking radiation seems not to play any important role in the case of large-sized black holes. In left-hand panel, Hawking temperature of Reissner-N\"{o}rdstrom black hole leaves the standard value earlier than the other two types of black holes. Next to it is Garfinkle-Horowitz-Strominger black hole. In the right-hand panel, Reissner-N\"{o}rdstrom and Garfinkle-Horowitz-Strominger are close to each other (dash-dotted curve moves to smaller values). The Hawking temperature of modified de Sitter-Schwarzschild black hole becomes larger than the standard value.

\begin{figure}[htb]
\centering{
\includegraphics[width=8cm,angle=0]{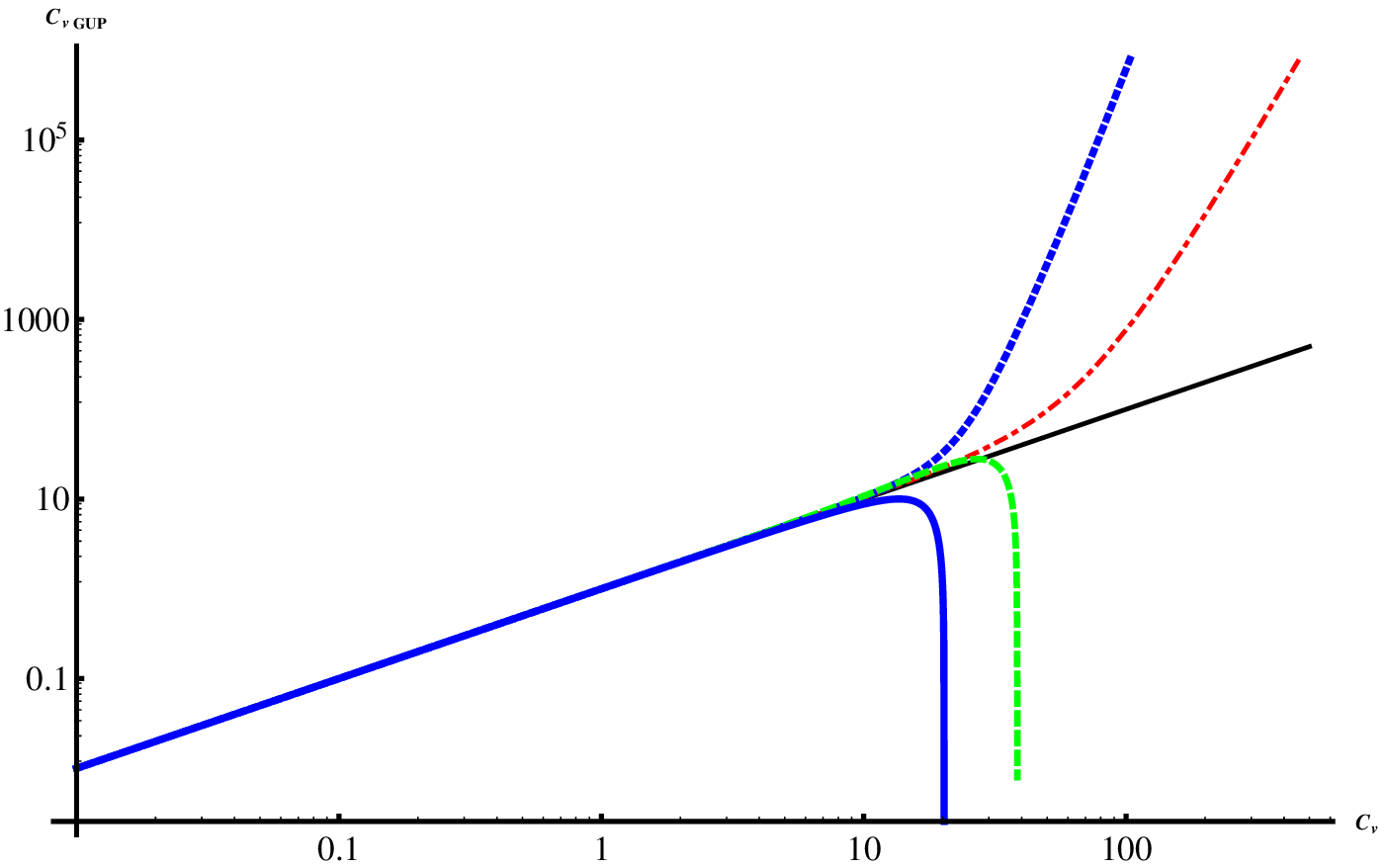}
\includegraphics[width=8cm,angle=0]{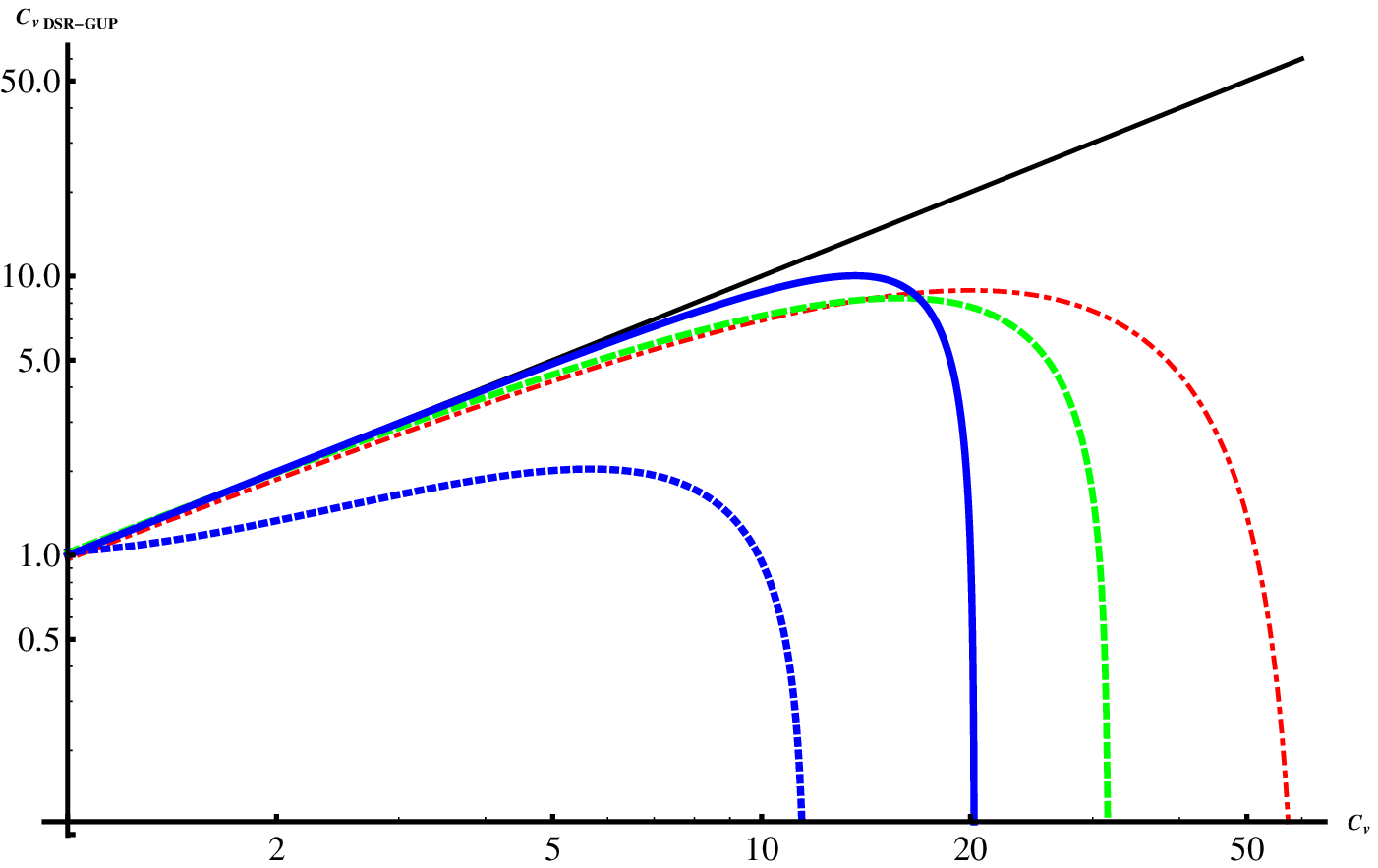}
\caption{The same as in Fig. \ref{fig:1} but for specific heat. 
\label{fig:4} }
}
\end{figure}

In Fig. (\ref{fig:4}), the modified specific heat is given in dependence on the standard one. The right-hand panel refers to the effect of linear GUP effect. In left-hand panel, we find that the modified specific heat of Garfinkle-Horowitz-Strominger black hole drops (energies) at small value of standard specific heat, while that of the other types exceed the standard value. Furthermore, standard de-Sitter-Schwarzshild black hole goes below the unmodified specific heat. Both modified de-Sitter-Schwarzshild and Reissner-N\"{o}rdstrom black holes increase. In right-hand panel, we find that the specific heat of all types of modified black holes and even standard de-Sitter-Schwarzshild black hole rapidly decrease with increasing standard specific heat.

\section{Conclusion}

We used various approaches for GUP and MDR in order to resolve the quantum geometrical corrections to the thermodynamical quantities of balck holes and the modified FLRW Universe according to correction to Bekenstein-Hawking entropy in four-dimensional black holes. For MDR, the sign of correction terms in both uncharged black holes and the charged one seems to depend on the quantum gravitational parameters $\alpha_{i}$, which is related to the particular model of the quantum gravity. There are considerable differences between the corrections due to GUP and MDR. For instance, the modified specific heat due to GUP and MDR vanishes at large standard specific heat. The correction due to GUP result has different behaviors. The specific heat of modified de-Sitter-Schwarzshild and Reissner-N\"{o}rdstrom black holes seems to increase at large values of the standard specific heat. In the earlier case, the black hole cannot exchange heat with the surrounding space. Thus, we predict existence of black hole remnants which may be considered as candidates for dark matter. In light of this, it would be appropriate to generalize the calculations in extra dimensions and investigate the possibilities of finding black holes in Large Hadron Collider and Ultra-High Energy Cosmic Rays, for instance.

The modification in the black hole entropy shed light on the differences between the consequences of the quantum gravity and that of the dispersion relation. In GUP and MDR approaches, the modified entropy of different black holes starts to exist at finite standard entropy, while in GUP approach, this is valid for modified de-Sitter-Schwarzshild and Reissner-N\"{o}rdstrom black holes. The modified entropy of Garfinkle-Horowitz-Strominger black hole starts from a finite value. Then, it decreases with increasing standard entropy. There exists a minimum value. Further increase in the standard entropy slowly brings the modified entropy to the reference line.

For the differences between the two approaches, we find that the modified energy density of Reissner-N\"{o}rdstrom black hole due to the linear GUP and MDR remains finite (smaller than the standard line). It disappears in MDR approach. Almost the same is valued for Garfinkle-Horowitz-Strominger black hole. In both approaches, the energy density of modified de Sitter-Schwarzschild remains finite.

The modified Hawking temperatures of different types of black holes are the same as the standard one, especially at low values of the  Hawking temperature.  In the quadratic GUP approach, the modified temperature drops at different values meaning that the Hawking radiation plays no important role in the case of large-sized black holes. Hawking temperature of Reissner-N\"{o}rdstrom black hole leaves the standard value earlier than the other two types. Next to it is Garfinkle-Horowitz-Strominger black hole. In linear GUP and MDR approaches, Reissner-N\"{o}rdstrom and Garfinkle-Horowitz-Strominger are close to each other (dash-dotted curve moves to smaller values). The Hawking temperature of modified de Sitter-Schwarzschild black hole becomes larger than the standard value.

A systematic investigation for the consequences of the three GUP and MDR approaches on the three metric types from various equations of states is planned in near future. We want to study the impacts of the different corrections on the evolution of cosmological parameters, such as scale factor, etc. In doing this, we might introduce possible modifications in Raychaudhuri equations, as well. 
 

\end{document}